\documentclass[aps,prx,twocolumn,superscriptaddress]{revtex4-1}
\usepackage{graphicx}
\usepackage{dcolumn}
\usepackage{color,soul}
\usepackage{soul}
\usepackage{amsfonts,amsmath,amssymb,bm}
\usepackage{xcolor}
\usepackage{hhline}
\usepackage[utf8]{inputenc}
\usepackage[colorlinks=true,allcolors=blue]{hyperref}
\usepackage{booktabs}
\usepackage{multirow}

\begin{document}
\title{Vacancy-related color centers in twodimensional silicon carbide monolayers}
\date{\today}

\author{M. Mohseni}
\affiliation{HUN-REN Wigner Research Centre for Physics, PO.\ Box 49, H-1525 Budapest, Hungary}
\affiliation{Lor\'and E\"otv\"os University, P\'azm\'any P\'eter S\'et\'any 1/A, H-1117 Budapest, Hungary}
\affiliation{Department of Physics, Isfahan University of Technology, Isfahan, 84156-83111, Iran}

\author{I. Abdolhosseini~Sarsari}\email{abdolhosseini@iut.ac.ir}
\affiliation{Department of Physics, Isfahan University of Technology, Isfahan, 84156-83111, Iran}
\author{S. Karbasizadeh}
\affiliation{Department of Physics, Isfahan University of Technology, Isfahan, 84156-83111, Iran}

\author{P\'eter Udvarhelyi}
\affiliation{HUN-REN Wigner Research Centre for Physics, PO.\ Box 49, H-1525 Budapest, Hungary}
\affiliation{Department of Atomic Physics, Institute of Physics, Budapest University of Technology and Economics, M\H{u}egyetem rakpart 3., H-1111 Budapest, Hungary}

\author{Q. Hassanzada}
\affiliation{The NOMAD Laboratory at Fritz Haber Institute of the Max Planck Society and IRIS Adlershof of the Humboldt University Berlin, Germany}

\author{T. Ala-Nissila}
\affiliation{QTF Centre of Excellence, Department of Applied Physics, Aalto University, FI-00076 Aalto, Finland}
\affiliation{Interdisciplinary Center for Mathematical Modelling and Department of Mathematical Sciences, Loughborough University, Loughborough, Leicestershire LE11 3TU, UK}

\author{A. Gali}\email{gali.adam@wigner.hu}
\affiliation{HUN-REN Wigner Research Centre for Physics, PO.\ Box 49, H-1525 Budapest, Hungary}
\affiliation{Department of Atomic Physics, Institute of Physics, Budapest University of Technology and Economics, M\H{u}egyetem rakpart 3., H-1111 Budapest, Hungary}
\affiliation{MTA-WFK Lend\"ulet "Momentum" Semiconductor Nanostructures Research Group}

\begin{abstract}
Basic vacancy defects in twodimensional silicon carbide (2D-SiC) are examined by means of density functional theory calculations to explore their magneto-optical properties as well as their potential in quantum technologies. In particular, the characteristic hyperfine tensors and optical excited states of carbon-vacancy, silicon-vacancy, and carbon antisite-vacancy pair defects in 2D-SiC are determined that are the key fingerprints of these defects that may be observed in electron paramagnetic resonance and photoluminescence experiments, respectively. Besides the fundamental characterization of the most basic native defects, we show that the negatively charged carbon antisite-vacancy defect is a promising candidate for realizing a near-infrared single-photon quantum emitter with spin doublet ground state, where the negative charge state may be provided by nitrogen doping of 2D-SiC. We find that the neutral carbon-vacancy with spin triplet ground state might be used for quantum sensing with a broad emission in the visible.
\end{abstract}

\maketitle
\newcommand{\RNum}[1]{\uppercase\expandafter{\romannumeral #1\relax}}
\section{Introduction}

Point defects in semiconductors and related materials may significantly alter the host material's electrical, optical, and magnetic properties. In traditional semiconductors, shallow-level point defects may be used as dopants to deliberately introduce carriers. In contrast, deep-level point defects act as carrier traps or recombination centers that can deteriorate the operation of semiconductor devices. Recently, this negative view about deep-level point defects has been radically changed after a room temperature electron spin resonance observation of a single point defect by optical means which was demonstrated for the negatively charged nitrogen-vacancy (NV) defect in diamond~\cite{Gruber1997, Doherty2013, Gali2019}. As the emission came from a single defect, it acted as a single-photon source or quantum emitter. The coherent manipulation, initialization, and readout of a single defect spin is a realization of a solid-state defect quantum bit or qubit that can be employed in quantum sensing, simulation, computation, and entanglement-based quantum communication studies and applications (e.g., Ref.~\onlinecite{Wolfowicz2021}). 

The success of the diamond NV center has motivated researchers to seek alternative paramagnetic quantum emitters that may have favorable magneto-optical or spin-coherence properties in technologically more mature materials than diamond, e.g., the threedimensional (3D) silicon carbide (SiC)~\cite{Gali2010, Weber2010, Gali2011, Koehl2011, Anderson2023}. Since then, single-photon emitters with or without coherent manipulation of electron spins have been found in various materials, including twodimensional (2D) ones~\cite{Zhang2020, Montblanch2023}. The platforms offered by 2D materials have certain crucial advantages as compared to their bulk counterparts in 3D materials: they typically have very high extraction efficiency with avoiding total internal reflection and can be well integrated with cavities and photonic waveguides and coupled to plasmonic structures~\cite{Aharonovich2016}.      

The first 2D material discovered~\cite{Novoselov666}, namely graphene, is a zero band gap material that cannot host single-photon emitters. Changing the chemical composition is one possible route to open the gap in carbon-based 2D materials. For example, every second carbon atom in graphene may be replaced by a silicon atom to produce a honeycomb-structured SiC material~\cite{doi:10.1021/acs.jpcc.5b04113, Susi2017} that we label by 2D-SiC in the context. This material should produce a band gap because of the partial polarization of the covalent bonds between carbon and silicon atoms and the breaking of the high symmetry of graphene as confirmed by first principles calculations at various levels of theory~\cite{Bekaroglu2010, Hsueh2011, Susi2017, PhysRevB.102.134103}.  

Very recently, the synthesis of this semiconductor material has been reported~\cite{Chabi2021, Polley2023}. Optical signals have been observed at room temperature for the 2D-SiC nanoflakes~\cite{Chabi2021}. In the absorption spectrum, two high intensity peaks were found with a maximum intensity at approximately 2.2 eV and 2.5~eV with a smaller peak at around 2.3~eV. In the photoluminescence spectrum, 1.0~eV-broad intense peak was found with a Gaussian shape where the maximum of the peak is located at around 2.58~eV~\cite{Chabi2021}. They associated this PL peak with the band edge emission based on the common appearance of intense peaks in the photoluminescence (excitation at 420~nm) and the absorption at around 2.6~eV in the same 2D-SiC nanoflakes sample. In a forthcoming experimental study, the 2D phase of SiC was found to be almost planar and stable at high temperatures up to 1200~$^\circ$C in vacuum~\cite{Polley2023}. The stability of 2D-SiC was previously predicted by density functional theory (DFT) calculations that showed all the phonon frequencies to be positive in the Brillouin zone (BZ) in its planar geometry~\cite{Bekaroglu2010}. As the existence and stability of this material have been recently demonstrated in experiments~\cite{Polley2023}, it can be considered as an interesting 2D platform to host quantum emitters and qubits. 

Recent advances in first principles techniques have made them highly predictive to explore potential defect qubit candidates~\cite{Gali2019, Zhang2020, Gali2023}. In our previous work, we carried out DFT calculations with a focus on Stone-Wales-related defects~\cite{PhysRevB.102.134103} to this end. On the other hand, many color centers in 3D and 2D materials are vacancy-like and can be created either during growth or by irradiation techniques~\cite{Zhang2020}. We note that the isolated single silicon-vacancy in 3D forms of SiC has been proven to be room temperature qubits~\cite{Riedel2012, Widmann2015, Zhang2020, soykal2016}, thus one might expect similar results in 2D-SiC. Furthermore, the magneto-optical characterization of the basic vacancy defects in 2D-SiC is fundamental to gain insight into the properties of real 2D-SiC materials. These issues motivated us for a detailed study of vacancy-type defects in 2D-SiC.  

\begin{figure*}[!htb]
\includegraphics*[scale=.075]{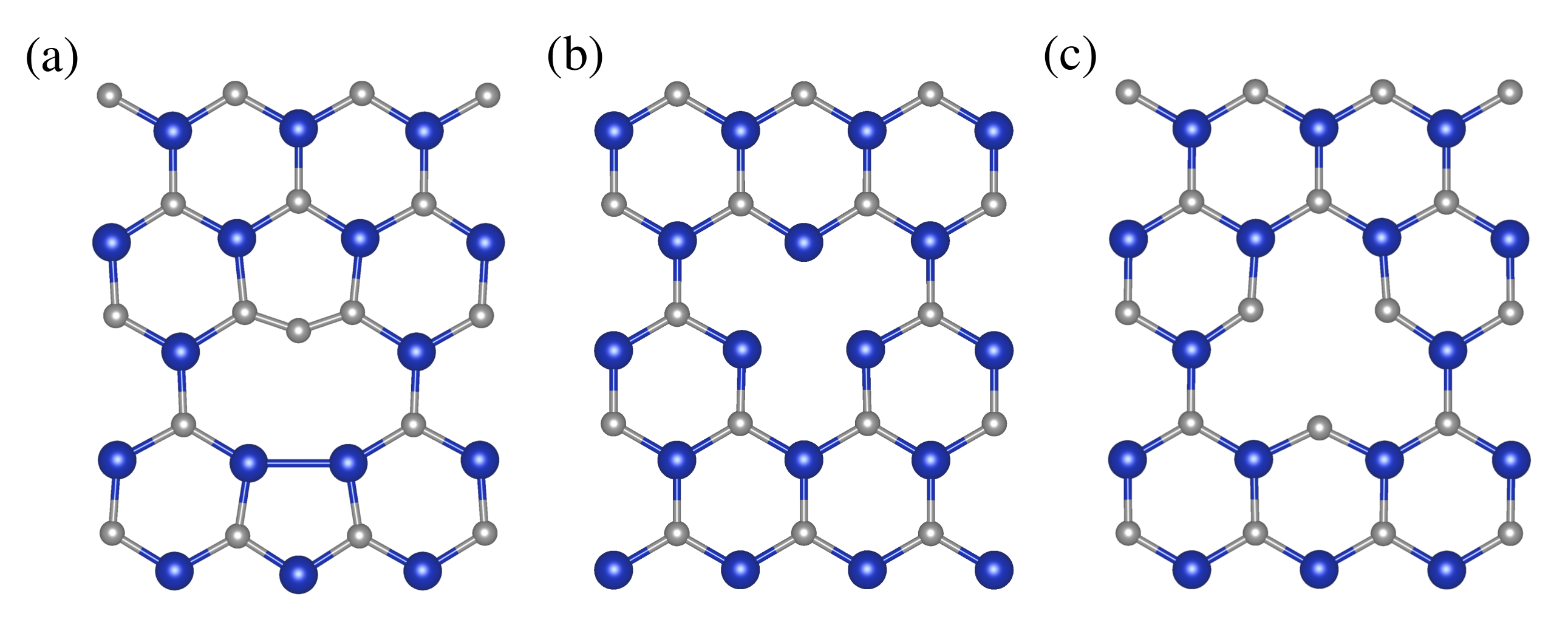}
\caption{
Atomic structures of (a) V$_{\text{C}}$-C$_{\text{Si}}$, (b) V$_{\text{C}}$ and (c) V$_{\text{Si}}$ defects after HSE06 geometry optimization within C$_{2v}$ point group symmetry for V$_{\text{C}}$-C$_{\text{Si}}$ and D$_{3h}$ for the other two defects. Si and C atoms are depicted by blue and gray spheres, respectively.}
\label{fig:defects}
\vspace{-3mm}
\end{figure*}

In 2D-SiC lattice, all the carbon and silicon atoms are equivalent by symmetry, thus the two simplest vacancy defects are the isolated carbon-vacancy (V$_{\text{C}}$) and silicon-vacancy (V$_{\text{Si}}$). We learnt from 3D-SiC~\cite{Rauls2000, Bockstedte2003, Umeda2007, Deak2011, Castelletto2014NatMat, Castelletto2014ACSNano} that V$_{\text{Si}}$ has another atomic configuration when one of the neighboring carbon atoms near the vacant site jumps to the vacancy by creating a carbon antisite-vacancy pair defect denoted by V$_{\text{C}}$-C$_{\text{Si}}$. The structures of these basic vacancy defects are depicted in Fig.~\ref{fig:defects}. The optical properties of these defects in part have been previously studied modeled in a small supercell~\cite{PhysRevB.102.134103} but a detailed magneto-optical characterization for these important defects is still lacking. Here, we apply advanced plane-wave supercell DFT calculations to compute the electronic structure, formation energies, photoionization thresholds, electron spin resonance hyperfine parameters, and the basic optical parameters of these defects. Furthermore, we simulate the photoluminescence spectrum for the most relevant defects. Our results provide a reference for future electron spin resonance and photoluminescence studies of 2D-SiC. We find that the negatively charged V$_{\text{C}}$-C$_{\text{Si}}$ defect is a good candidate to operate as a near-infrared quantum emitter and may be used to realize a qubit in 2D-SiC. Furthermore, the carbon-vacancy has a triplet ground state with a broad emission in the visible that might be harnessed as a quantum sensor if optically detected magnetic resonance is confirmed for the defect.

\begin{figure*}[t]
\includegraphics*[scale=.34]{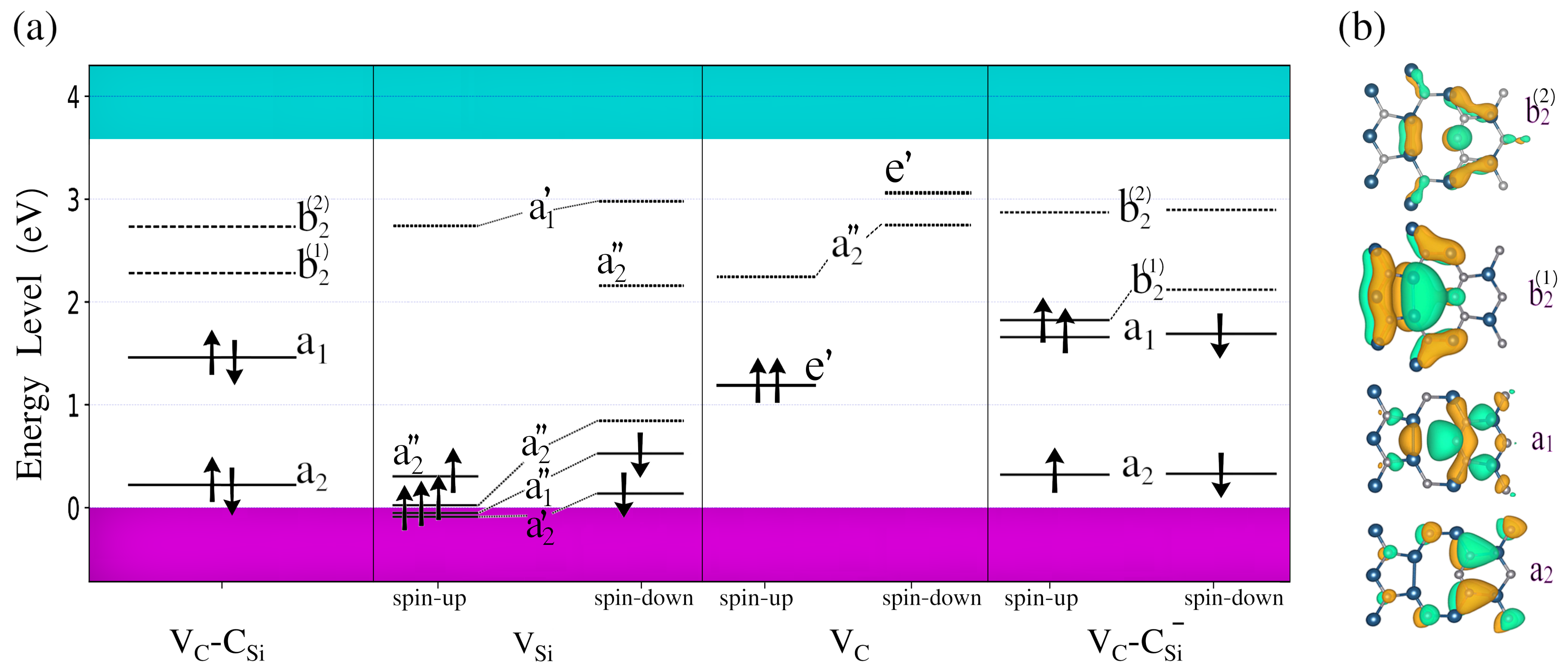}
\caption{
Illustration of the electronic structure (a) in the ground state of V$_{\text{C}}$-C$_{\text{Si}}$, V$_{\text{Si}}$, V$_{\text{C}}$ and V$_{\text{C}}$-C$_{\text{Si}}^-$ defects (from left to right, respectively) as obtained by HSE06 calculations. Arrows show the spin state of the electron. Each electronic state is labeled according to the irreducible representation of the orbitals. Conduction band minimum (CBM) and valence band maximum (VBM) are shown with cyan and purple colors, respectively. (b) The respective Kohn-Sham wavefunctions of the electronic states of V$_{\text{C}}$-C$_{\text{Si}}^{-}$. Light green and brown lobes exhibit negative and positive isovalues. The isosurface absolute value is $10^{-8}$~\AA$^{-3}$.}
\label{fig:elstr}
\vspace{-3mm}
\end{figure*}

\section{Method} \label{sec:method}
In this work, we employ plane-wave supercell DFT calculations~\cite{PhysRev.140.A1133,PhysRev.136.B864} with the projector-augmented-wave (PAW) method~\cite{PhysRevB.50.17953,PhysRevB.59.1758} as implemented in the Vienna Ab-initio Simulation Package (VASP)~\cite{PhysRevB.54.11169,Kresse1996}. The exchange-correlation functional of Heyd, Scuseria, and Ernzerhof~\cite{doi:10.1063/1.1564060} was used with its parameters set to standard values (HSE06). For all unit cell calculations, a dense $\Gamma$-centered $k$-point mesh of $9 \times 9 \times 1$ was used, while supercell calculations were performed using a single $\Gamma$ point. The geometry was optimized with the criterion of 10$^{-2}$~eV/\AA{} per atom for the Hellman-Feynman forces and a kinetic energy cutoff of 450~eV was used. We employed a vacuum region of 12~\AA{} and 24~\AA{} in the $z$ direction for the $6 \times 6 \times 1$ and $8 \times 8 \times 1$ supercells, respectively, minimizing all interactions between periodic images of the embedded defects. 
The $\Delta$SCF method~\cite{Gali2009} was implemented in VASP to calculate the electronic excited state with optimizing the geometry in the excited state where all the atoms are allowed to relax in the geometry optimization procedure. After geometry optimization the zero-phonon-line energies can be calculated and the obtained geometry in the electronic excited state also contributes to the simulation of the phonon sideband of the photoluminescence spectrum as will be explained below.

The equilibrium geometry in the electronic ground and excited states may differ. The difference in the geometries of the electronic ground and excited states ($\Delta Q$) may be defined~\cite{PhysRevLett.109.267401} as
\begin{equation}
\label{eq:dQ}
\Delta Q^{2}=\sum_{\alpha, i} m_{i } (R^{(\rm e)}_{\alpha,i} -R^{(\rm g)}_{\alpha,i})^2\text{,} 
\end{equation}
where $i$ indicates the atom index, $\alpha=(x,y,z)$, $m_{i }$ is the atomic mass of species $i$ and $R^{({\rm g/e})}_{\alpha,i}$ is the position of atom $i$ in the ground ($g$)/excited ($e$) state. Usually, the larger the difference the larger the phonon sideband in the photoluminescence spectrum for the optically allowed transition.

For the calculation of phonon frequencies and normal coordinates, we used the same parameters as above for the vacuum region, cutoff energy, and $k$-point sampling. However, the more traditional functional of Perdew, Burke, and Ernzerhof (PBE)~\cite{PhysRevLett.77.3865} was applied to optimize the forces on each atom within a threshold of 10$^{-3}$~eV/\AA{}. The phonons were calculated in the electronic ground state by DFT perturbation theory as implemented in VASP. Choosing the exchange-correlation functional of PBE over HSE06 in our phonon calculations provides reasonably accurate results while significantly reducing the computational time.
   
The photoluminescence spectrum of the emitters may reveal their potential in various quantum technology applications. In quantum communication applications, the coherent emission only comes at the no-phonon or zero-phonon-line (ZPL) emission. The fraction of ZPL and total emission can be observed in experiments from the Debye-Waller factor ($W$). In the theoretical simulation of the emitters, the Huang-Rhys (HR) theory may be applied to calculate the phonon sideband of the photoluminescence spectrum~\cite{Alkauskas_2014, Thiering2018} which requires the optimized geometries in the ground and optical excited states as well as the phonons in the ground state. The total Huang-Rhys factor ($\mathcal{S}$) indicates in the HR theory how many effective phonons participate in the optical transition and it is related to $W$ by
\begin{equation}
W=e^{-\mathcal{S}}\text{.}
\end{equation}
To calculate the phonon sideband of the spectrum, the partial HR factors should be calculated as explained in Ref.~\onlinecite{Alkauskas_2014}. The partial Huang-Rhys factor is $S(\hbar\omega)=\frac{\omega {Q_\omega}^2}{2 \hbar}$ where $Q_\omega$ is defined as $\Delta Q$ in Eq.~\eqref{eq:dQ} but $R^{(\rm e)}_{\alpha, i}$ should be replaced by the coordinates of ions associated with the normal coordinate of the $\omega$ phonon. The temperature-dependent spectrum can be calculated by the Fourier transform of the partial HR factor as
\begin{equation}
S(t)=\int_{0}^{\infty}S(\hbar \omega)F(\hbar \omega)e^{-i\omega t} d(\hbar \omega)\text{,}
\end{equation}
where $F(\hbar \omega)$ is the Bose-Einstein distribution function,
\begin{equation}
F(\hbar \omega)= \frac{1}{e^{{\hbar w}/{k_{\rm B}T}}-1}\text{,}
\end{equation}
where $k_{\rm B}$ is the Boltzmann-constant and $T$ is temperature. 

The phonon sideband of the spectrum can then be defined as
\begin{equation}
L(E_{\rm ZPL}-\hbar \omega)=\frac{C\omega^{3}}{2\pi}\int_{-\infty}^{\infty} e^{[S(t)-S(0)]} e^{i\omega t-\gamma\left |t  \right |} dt\text{,}
\end{equation}
where $C$ is a normalization constant and $\gamma$ is the broadening of the ZPL emission. We apply HR theory to calculate the HR factors and the shape of the phonon sideband of the photoluminescence spectrum. We note that we applied HSE06 optimized geometries and PBE phonons in this procedure using an in-house code. This approach produced an accurate optical spectrum for defects in diamond and 3D-SiC~\cite{thiering2017, Thiering2018, Csore2022, Csore2022PRB}. We further note that the temperature broadening of the ZPL emission or complicated anharmonicity and renormalization related effects are not considered in the simulation of temperature broadening in the optical spectrum which goes beyond the scope of our study.

For the paramagnetic defects, we determined the key magnetic parameters that may be observed by electron spin resonance techniques. For ensemble defect measurements, the statistics of the nuclear spin distribution proximate to the defect and the electron spin density distribution will produce a unique fingerprint caused by the hyperfine interaction. Therefore, the electron paramagnetic resonance (EPR) spectrum is a powerful method to identify the atomic structure of the defect, in particular, when combined with the DFT simulated spectra. The shape of the EPR spectrum is mostly set by the hyperfine (HF) tensor of the nuclear spins which is given by
\begin{equation}
\label{eq:HF}
A^{(I)}_{ij}=\frac{1}{2S}\int d^3r n_s(r) \gamma_I\gamma_e \hbar^2[(\frac{8\pi}{3}\delta(r))+(\frac{3r_ir_j}{|r|^5}-\frac{\delta_{ij}}{|r|^3})]\text{,}
\end{equation}
where $n_s(r)$ is the spin density of spin state $S$ at the site $r$, $\gamma_e$ is the electron Bohr magneton and $\gamma_I$ is the nuclear Bohr magneton for nucleus $I$, whereas $r_i$ refers here to the unit vector of $i={x,y,z}$. The first and second terms in the parentheses represent the Fermi-contact and dipole-dipole terms, respectively. The dominance of the Fermi-contact term over the dipole-dipole one will produce a characteristic hyperfine sideband in the EPR spectrum as the Fermi-contact term strongly depends on the spin density localized at the nucleus. We note that Eq.~\eqref{eq:HF} is modified within the PAW formalism~\cite{Blochl2000}, and the spin polarization of the core orbitals is taken into account for accurate HSE06 calculations~\cite{Szasz2013}. 

The EPR spectrum of high-spin ($S > 1/2$) defects may be further modified even in the absence of an external magnetic field, due to the electron spin-spin dipole interaction. The so-called zero-field-splitting (ZFS) can be described by the Hamiltonian operator
\begin{equation}
\label{eq:Hssij}
\hat{H}_{\rm SS,ij}=-\frac{\mu_{\rm 0}}{4\pi}g_e^2\beta^2 \sum_{i>j}
\left[\frac{\hat{S}_i\hat{S}_j}{|\mathbf{r}_{ij}|^3}-\frac{3(\hat{S}_i\mathbf{r}_{ij})(\mathbf{r}_{ij}\hat{S}_j)}{|\mathbf{r}_{ij}|^5}\right] \text{,}
 \end{equation}
%
%
where $\mathbf{r}_{ij}=\mathbf{r}_i - \mathbf{r}_j$ is the vector between the spins, $\hat{S}_{i}$ is the spin operator vector, $\beta$ is the Bohr magnet on of the electron, and $\mu_{0}$ is the magnetic permeability of vacuum. The formula in Eq.~\eqref{eq:Hssij} can be cast to
\begin{equation}
\hat{H}_{\rm SS, ZFS}=\Sigma_{i>j}{\bf S}_{i} D_{ij} {\bf S}_{j}\text{,}
\end{equation}
where {\bf S} is the total spin vector and {\bf D} is zero-ﬁeld-splitting tensor (ZFST). Since the eigenvalue framework matrix is diagonal, the spin-spin Hamiltonian can be expressed as
\begin{align}
\hat{H}_{\rm SS, ZFS}&=D_{ xx}S_{x}^2+D_{yy}S_{y}^2+D_{zz}S_{z}^2\\
&=D\left(S_{z}^2-\frac{S(S+1)}{3} \right )+ \frac{E(S_{\rm +}^2+S_{\rm -}^2)}{2}
\text{,}
\end{align}
where $D_{ij}$ are the components of the ZFST, $S_{\alpha}$ with $\alpha = x,y,z$ are the spin matrices, ${S}_{\pm}={S}_{x} \pm i{S}_{y}$ are the spin raising and lowering operators, $S^2=S_{x}^2+S_{y}^2+S_{z}^2$, and $D$ and $E$ are the two parameters of the ZFST in the eigenvalue framework. The parameters $D$ and $E$ can be expressed as
\begin{equation}
\label{eq:DE}
D=\frac{3}{2}D_{zz} \text{ and } E=\frac{D_{yy}-D_{xx}}{2}\text{,}
\end{equation}
where $D$ is the axial parameter whereas $E$ is the orthorhombic parameter, so the latter is non-zero for defects with low symmetries. We consider the electron spin dipole-dipole interaction as a source of ZFS which is calculated as was implemented by Martijn Marsman in VASP within the PAW formalism~\cite{Rayson2008, Bodrog_2014, Viktor2018}.

As we will note below, a strongly correlated electronic structure could develop in certain cases which is manifested as symmetry breaking solution for the Kohn-Sham wavefunctions in the spinpolarized hybrid DFT calculations. This solution may yield a proper total energy of the system but it is improper for the spin density or spin density matrix related quantities. To approximate the HF tensor and ZFST for these cases, we constrained the Kohn-Sham wavefunctions to reflect the symmetry of the system. This procedure was successfully applied to oxygen defects in diamond~\cite{gali2016}.

\section{Results} \label{sec:results}

We investigate vacancy defects in a self-standing monolayer of 2D-SiC. We first briefly describe the host 2D-SiC system and then turn to the basic electronic structure of the vacancy defects. We define the formation energies of the defects which reveal their relative stability,
charge transition levels, and photoionization thresholds. We then determine the ground-state spin properties of the paramagnetic vacancy defects. Finally, we compute and discuss their optical properties in detail by providing the photoluminescence spectrum for the most relevant vacancy defects. In the discussion, we also show results for the n-type doping of 2D-SiC in the context.

The pristine 2D-SiC crystal is optimized using HSE06 which yields a lattice constant of $a = 3.07$~\AA{} and Si-C bond length of 1.77~\AA{}, which agree well with the experimental data at 3.1~\AA{} and 1.79~\AA{} (Ref.~\onlinecite{Chabi2021}), respectively. The calculated band gap is 3.58~eV, which broadly agrees with previous calculations at various levels of theory~\cite{PhysRevB.102.134103, Hsueh2011}. After optimizing the local 2D-SiC structure, we construct monolayer supercells and embed the three vacancy defects as depicted in Fig.~\ref{fig:defects}.

\begin{figure*}[t]
\centering
\includegraphics*[scale=0.11]{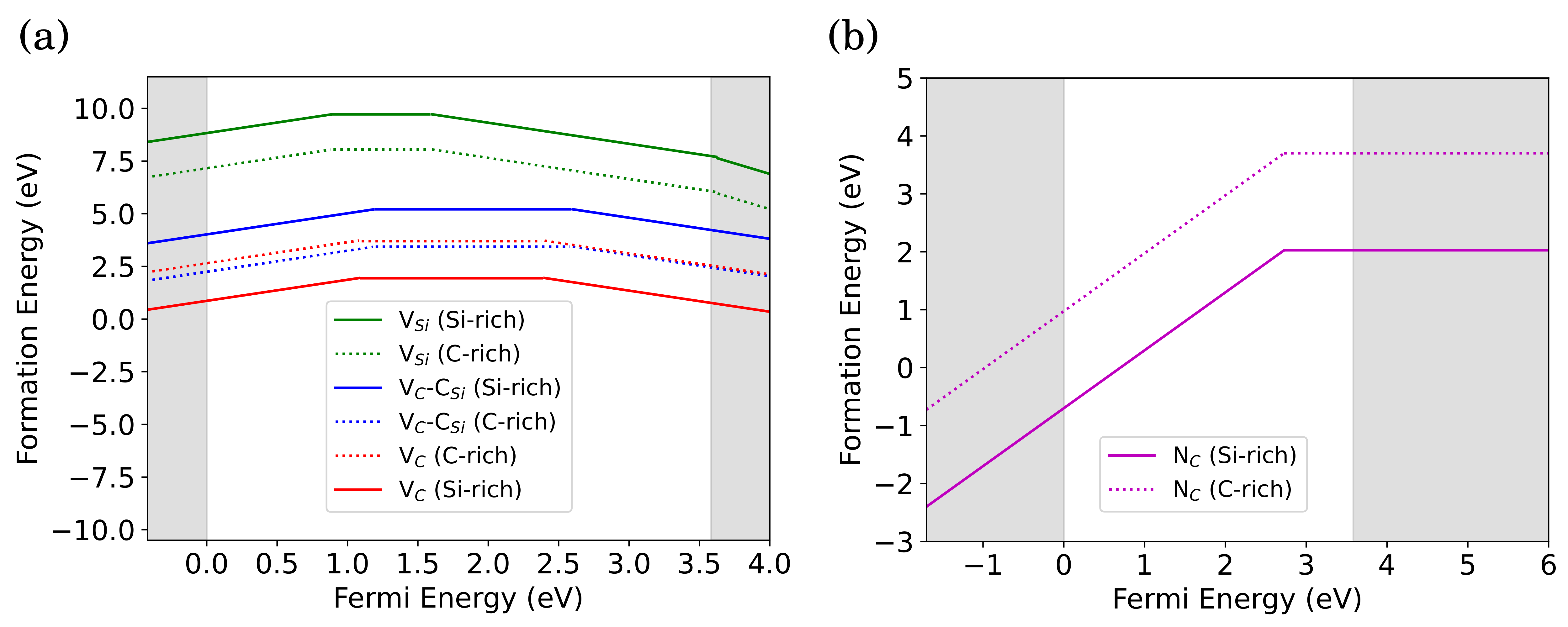}
\caption{
HSE06 computed the formation energy as a function of the Fermi-level for (a) the vacancy defects V$_{\text{C}}$-C$_{\text{Si}}$, V$_{\text{C}}$ and V$_{\text{Si}}$, and (b) N$_{\text{C}}$ in Si-rich and C-rich growth conditions. The valence band maximum is aligned to zero, and the band regions are shown by shaded areas.} 
\label{fig:formen}
\end{figure*}
%
%
\begin{figure*}[t]
\centering
\includegraphics*[scale=0.43]{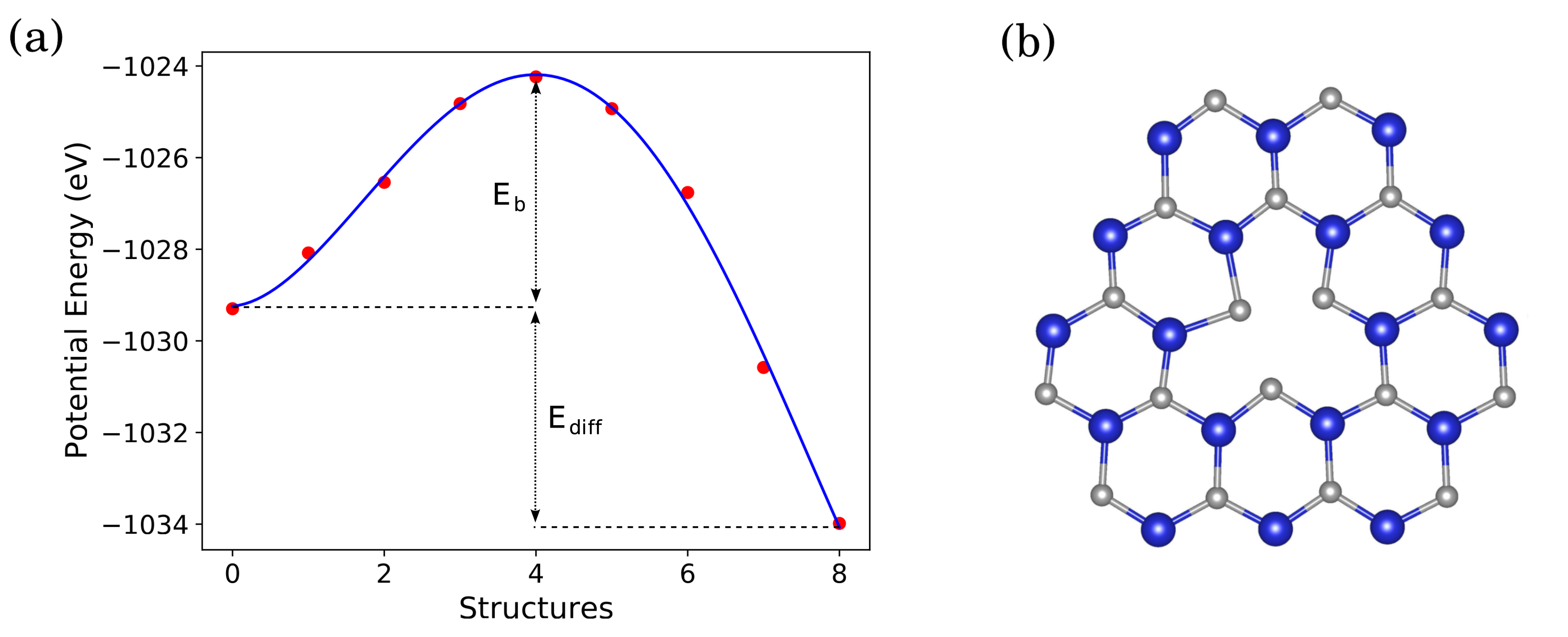}
\caption{(a) The nudged elastic band HSE06 calculation from V$_{\text{Si}}$ (structure No.\ 0) to V$_{\text{C}}$-C$_{\text{Si}}$ (structure No.\ 8), where E$_{\text{b}}$ and E$_{\text{diff}}$ label the barrier energy between the two Si vacancies and the adiabatic potential energy difference between two V$_{\text{Si}}$ and V$_{\text{C}}$-C$_{\text{Si}}$ defects, respectively. (b) The geometry of the saddle point at structure No.\ 4. Si and C atoms are depicted by blue and gray spheres, respectively.}
\label{fig:nudgplot}
\end{figure*}

\subsection{Electronic structure of the vacancy defects}

The neutral vacancy defects have dangling bonds that introduce occupied and empty defect levels in the fundamental band gap as depicted in Fig.~\ref{fig:elstr} and labeled by the irreducible representation of the associated point group as given in Fig.~\ref{fig:defects}. We illustrate the symmetry of the Kohn-Sham wave functions on the V$_{\text{C}}$-C$_{\text{Si}}$ defect in Fig.~\ref{fig:elstr}(b). The V$_{\text{C}}$ and V$_{\text{Si}}$ defects have $S=1$ high-spin ground state with D$_{3h}$ symmetry in the spinpolarized HSE06 calculations. We note that the spinpolarized hybrid DFT solution exhibits $C_{2v}$ symmetry for the Kohn-Sham wavefunction of V$_{\text{Si}}$ which is the signature of highly correlated state and the possibility that the true ground state of the system could be a highly correlated singlet state, similar to the neutral silicon-vacancy in 3D-SiC~\cite{Deak1999}. This means that the total energy of this state is less accurately calculated than the non-highly correlated ones, and the true spin state of the electronic ground state is uncertain. Nevertheless, the triplet state may be observed at elevated temperatures, thus we further consider this high-spin state for V$_{\text{Si}}$.  

We find that all the considered defects contain filled and empty levels in the gap. This implies that all the defects may be ionized by varying the position of the Fermi-level in the band gap. To determine the stability of the ionized defects the formation energy of the defects is computed which also reveals the relative stability of these defects in thermal equilibrium. In the next section, the details about the calculations of the formation energy of defects are described which is followed by the HSE06 results.

\subsection{Formation energies, relative stability, charge transition levels, doping, and photoionization thresholds} \label{ssec:formen}

The stability of the defect structures is determined via the formation energy. For a defect in charge state $q$, the formation energy ($E^{\rm f}$) is given by~\cite{Zhang1991, PhysRevB.86.045112},
\begin{align}\label{eq:formen}
E_{q}^{\rm F}(\epsilon _{\rm F}) = (&E_{q}^{\rm tot}+E_{q}^{\rm corr})-E_{\rm perfect}-n_{x}\mu _{x}\\
& +q(\epsilon_{\rm VBM}^{\rm perfect} +\epsilon_{\rm F} -\Delta V) \nonumber \text{,}
\end{align}
which depends on the Fermi-level energy $\epsilon _{\rm F }$. The energies are referenced to the calculated valence band maximum (VBM) $\epsilon_{\rm VBM}^{\rm perfect}$ of the perfect supercell. The quantities $E_{q}^{\rm tot}$ and $E_{q}^{\rm corr}$ are the total energy of the defected system and the finite-size electrostatic correction, respectively, both representing the charge state $q$. $\Delta V$ is the alignment between the electrostatic potentials of the pristine and defective supercells. The term $E_{\rm perfect}$ is the total energy of a pristine supercell of the same size. The parameter $n_{x}$ refers to the number of atoms of type $x$ added (positive) or removed (negative) from the perfect crystal with the appropriate chemical potential $\mu_{x}$. 

The electrostatic corrections are calculated within the spirit of Freysoldt-Neugebauer-van de Walle (FNV) method~\cite{Freysoldt} as implemented in the software package CoFFEE~\cite{NAIK2018114} which goes as
\begin{align}\label{eq:Ecorr}
E_{q}^{\rm corr} = &E_{q}^{\rm lat}-q\Delta V_{\rm q-0/m}\text{,}
\end{align}
where $E_{q}^{\rm lat}= E_{q}^{\rm iso, m}-E_{q}^{\rm per, m}$ is the long-range interaction energy
which is obtained by solving the Poisson equation with using a model charge distribution $\rho^{\rm m}{(r)}$ and a model dielectric profile $V_{q}^{\text{per}, \rm m}{(r)}$. 
$E_{q}^{\text{per}, \rm m}$ is given by
\begin{align}\label{eq:Ecorr2}
E_{q}^{\text{per}, \rm m} = \frac{1}{2} \int_{\Omega}\rho^{\rm m}{(r)}V_{q}^{\text{per}, \rm m}{(r)}dr^3 \text{,}
\end{align}
where the integral is over the supercell volume $\Omega$. The quantity $E_{q}^{\text{per}, \rm m}$ is evaluated for larger supercells and extrapolated to obtain $E_{q}^{\text{iso}, \rm m}$.
$\Delta V_{q-0/{\rm m}}$ gives the DFT difference electrostatic potential to the model potential as
\begin{align}\label{eq:deltaV}
\Delta V_{q-0/{\rm m}} = (&V_{q}^{\text{DFT}}- V_{0}^{\text{DFT}})|_{\text{far}}-V_{q}^{\text{per},\rm m}|_{\text{far}}\text{,}
\end{align}
where the planar-averaged electrostatic potentials are obtained along the $z$ direction.
We apply this charge correction scheme to the special excited state of the neutral defects which can be described by defect level to band edge transition in order to account for the supercell size correction of this special excited state (see Refs.~\onlinecite{Zhang2020PRL, Udvarhelyi2022, Gali2023}). The charge correction in the total energy of V$_{\text{C}}$-C$_{\text{Si}}$, V$_{\text{Si}}$ and V$_{\text{C}}$ defects in their positive and negative charge states is around $0.6$~eV in our supercell. 

When computing the formation energy, different growth conditions have to be taken into account. 
The two extreme growth conditions for the 2D-SiC binary compound correspond to Si-rich and C-rich ones.
When including these conditions into the formation energy, the chemical potential $\mu_{x}$ must be evaluated under such conditions. In the Si-rich case, $\mu_{\rm Si}$ is calculated from a silicon crystal. The term $\mu_{\rm C}$ is then taken from the equilibrium condition
\begin{equation}\label{eq:muSiC}
\mu_{\rm SiC}=\mu_{\rm Si}+\mu_{\rm C}\text{,}
\end{equation}
where $\mu_{\rm SiC}$ is the chemical potential of the 2D-SiC primitive cell. In C-rich condition, $\mu_{\rm C}$ is taken from the diamond crystal and $\mu_{\rm Si}$ is determined again with Eq.~\eqref{eq:muSiC}. The formation energies of V$_{\text{C}}$-C$_{\text{Si}}$, V$_{\text{Si}}$-C$_{\text{Si}}$, and V$_{\text{Si}}$ are then computed using Eq.~\eqref{eq:formen}. 

The results are plotted in Fig.~\ref{fig:formen}(a) for both Si-rich and C-rich conditions. The crossing points in the plot for a given defect show the charge transition levels that are listed in Table~\ref{tab:CTL}. The $(+|0)$ and $(0|-)$ charge transition levels for the V$_{\text{C}}$-C$_{\text{Si}}$, V$_{\text{Si}}$ and V$_{\text{C}}$ defects in Fig.~\ref{fig:formen}(a) appears deep in the fundamental band gap. In particular, the $(+|0)$ and $(0|-)$ charge transition levels appear at $\epsilon_{\rm VBM}^{\rm perfect}+1.2$~eV and $\epsilon_{\rm VBM}^{\rm perfect}+2.6$~eV for V$_{\text{C}}$-C$_{\text{Si}}$, $\epsilon_{\rm VBM}^{\rm perfect}+0.9$~eV and $\epsilon_{\rm VBM}^{\rm perfect}+1.6$~eV for V$_{\text{Si}}$, and $\epsilon_{\rm VBM}^{\rm perfect}+1.1$~eV and $\epsilon_{\rm VBM}^{\rm perfect}+2.4$~eV for V$_{\text{C}}$, respectively. We checked that the double negative charge state is unstable for V$_{\text{C}}$-C$_{\text{Si}}$, V$_{\text{Si}}$ and V$_{\text{C}}$ defects.


We find that the V$_{\text{C}}$ defect is the most stable basic vacancy-type defects in 2D-SiC which is followed by the V$_{\text{C}}$-C$_{\text{Si}}$ defect in hierarchy. Our calculations imply that V$_{\text{Si}}$ is significantly less stable than its twin counterpart V$_{\text{C}}$-C$_{\text{Si}}$, irrespective to the position of the Fermi-level. This might indicate that V$_{\text{Si}}$ does not occur in 2D-SiC. However, V$_{\text{Si}}$ might form by irradiation techniques that can further survive as a metastable defect. Indeed, we find that the barrier energy of the transformation from V$_{\text{Si}}$ to V$_{\text{C}}$-C$_{\text{Si}}$ is 4.96~eV (see Fig.~\ref{fig:nudgplot}) as obtained by nudged elastic band method~\cite{Henkelman2000} which is very high and should be stable even at high-temperature annealing. As a consequence, the properties V$_{\text{Si}}$ defects are further considered in the context. We further note that if the V$_{\text{C}}$ defect becomes mobile at a certain annealing temperature then it may be combined with an existing C$_{\text{Si}}$ defect which has relatively low formation energy at 4.43~eV in Si-rich condition. In that case, additional V$_{\text{C}}$-C$_{\text{Si}}$ defects may be created by complex formation. We find that the complex formation is favored by $1.16$~eV in the neutral charge state (when the Fermi-level is pinned to the middle of the gap). 

In order to stabilize the acceptor states of the vacancy defects in thermal equilibrium, donors should be introduced into 2D-SiC. It has been shown in SiC nanotubes~\cite{Gali2006} that nitrogen-substituting carbon behaves as a donor. SiC nanotubes can be considered as tubular forms of SiC sheets, thus this idea may be transferable to 2D-SiC. We considered nitrogen substituting carbon, N$_{\text{C}}$, as a donor candidate in 2D-SiC. In order to calculate its formation energy, the chemical potential of nitrogen should be known in Eq.~\eqref{eq:formen}. To this end, we considered the most stable structure of the hexagonal Si$_{3}$N$_{4}$ crystal with a $P6_{3}/m$ space group. Si$_{3}$N$_{4}$ acts as the second phase that can determine the solubility limit. It is chosen because it is the most stable structure that can form between nitrogen and either silicon or carbon, hence giving us the limits for when nitrogen is introduced to the system. Thus, the chemical potential of nitrogen ($\mu_{\rm N}$) can be defined as
\begin{equation}\label{eq:muN}
\mu_{\rm Si_{3} N_{4}}=3 \mu_{\rm Si}+4 \mu_{\rm N}\text{,}
\end{equation}
where $\mu_{\rm Si_{3} N_{4} }$ is the chemical potential of the hexagonal Si$_{3}$N$_{4}$ primitive cell and $\mu_{\rm Si}$ can vary between the Si-rich and C-rich conditions as set above. We find that N$_{\rm C}$ is, in fact, a donor in 2D-SiC with $(+|0)$ level at $\epsilon_{\rm VBM}^{\rm perfect}+2.7$~eV or $\epsilon_{\rm CBM}^{\rm perfect}-0.88$~eV (where CBM refers to a conduction band minimum) as shown in Fig.~\ref{fig:formen}(b). The formation energy of N$_\text{C}$ is low; thus it is likely that N$_\text{C}$ doped 2D-SiC can support negatively charged vacancy defects.
\begin{table}[t]
\setlength\extrarowheight{6pt}
\setlength\tabcolsep{1pt}
\caption{\label{tab:CTL}
Calculated charge transition level ($\varepsilon$) of the considered defects at HSE06 level, e.g., $\varepsilon(+|0)$ is the position of the Fermi-level for which transition from positive to neutral charge states occurs. The levels are aligned to the valence band maximum and the calculated band gap is 3.58~eV. The unstable levels are shown in gray ink.}
\begin{ruledtabular}
\begin{tabular}{cccc}
Defect & $\varepsilon(+|0)$ (eV) &  $\varepsilon(0|-)$ (eV)& $\varepsilon(-|2-)$ (eV) \\ 
\hline
V$_{\text{Si}}$&0.9&1.6&\textcolor{gray}{3.6}\\
V$_{\text{C}}$-C$_{\text{Si}}$&1.2&2.6&\textcolor{gray}{4.8}\\
V$_{\text{C}}$&1.1&2.4&\textcolor{gray}{4.9}\\
N$_{\text{C}}$&2.7& & \\
\end{tabular}
\end{ruledtabular}
\end{table}

Another way to realize a given charge state of the defect is photoionization upon illumination which is not a thermal equilibrium process. By laser irradiation with energy above the ionization threshold, a defect can be driven to a photoionized charge state. Although, photoionization may be harmful to the realization of a stable emission from a given defect when it drives out the defect from its target charge state, leading to blinking or permanent quenching. The photoionization threshold energies may be lower than the neutral excitation energies of the defect in a given charge state, so then photoionization can be avoided, at least, in the regime of linear optical excitation. The photoionization threshold energies of the vacancy defects can be read out in Fig.~\ref{fig:formen}(a). We illustrate the role of photoionization in the relevant V$_{\text{C}}$-C$_{\text{Si}}$ defect with $(0|-)$ level at $\epsilon_{\rm VBM}^{\rm perfect}+2.6$~eV. Assuming that the defect was in the negative charge state, an electron may be promoted from the occupied defect level to the conduction band by a $(3.58-2.6)=0.98$~eV excitation energy. Thus, the condition of the existing photoluminescence signal from the negatively charged V$_{\text{C}}$-C$_{\text{Si}}$ is that the associated optical excitation energies should be smaller than 0.98~eV. Using the same argument, these values are 1.98~eV for the single negative charge state of V$_{\text{Si}}$ and 1.18~eV for the single negative charge state of V$_{\text{C}}$, respectively. For the complete picture, one has to consider the deep donor levels of the defects, too. If the defects were in their neutral charge state then the value of $(+|0)$ occupation level with respect to the conduction band minimum provides photoionization thresholds when the electron is promoted from the in-gap defect level to the conduction band edge. As an example, the neutral carbon-vacancy, V$_\text{C}^{0}$, has such photoionization threshold at $(3.58-1.1)=2.48$~eV. However, there is a lower ionization threshold at around 2.40~eV which transforms V$_\text{C}^{0}$ to V$_\text{C}^{-}$. This is a typical amphoteric defect where both the electrons and holes may be generated by illumination or captured in the dark by the defect. However, once we arrive at the negative charge state by photoexcitation with laser energy higher than 2.4~eV then the next photon will turn V$_\text{C}^{-}$ to V$_\text{C}^{0}$ because the ionization threshold of this process is at around $(3.58-2.4)=1.18$~eV. If the laser energy is higher than 2.48~eV then the defect may be ionized to V$_\text{C}^{+}$. However, the next photon would bring it back again to V$_\text{C}^{0}$ because the ionization threshold of this process is just 1.1~eV. The conclusion is that the defect dynamically stays in the neutral charge state upon laser illumination even at high-energy illumination ($>2.4$~eV) and it definitely stays there when the excitation energy is just below the $(0)\rightarrow (-)$ photoionization threshold energy. A similar scenario can be drawn for the neutral V$_{\text{C}}$-C$_{\text{Si}}$ too.

\subsection{Electron spin resonance parameters}

\begin{figure*}[t]
\centering
\includegraphics*[width=1\textwidth]{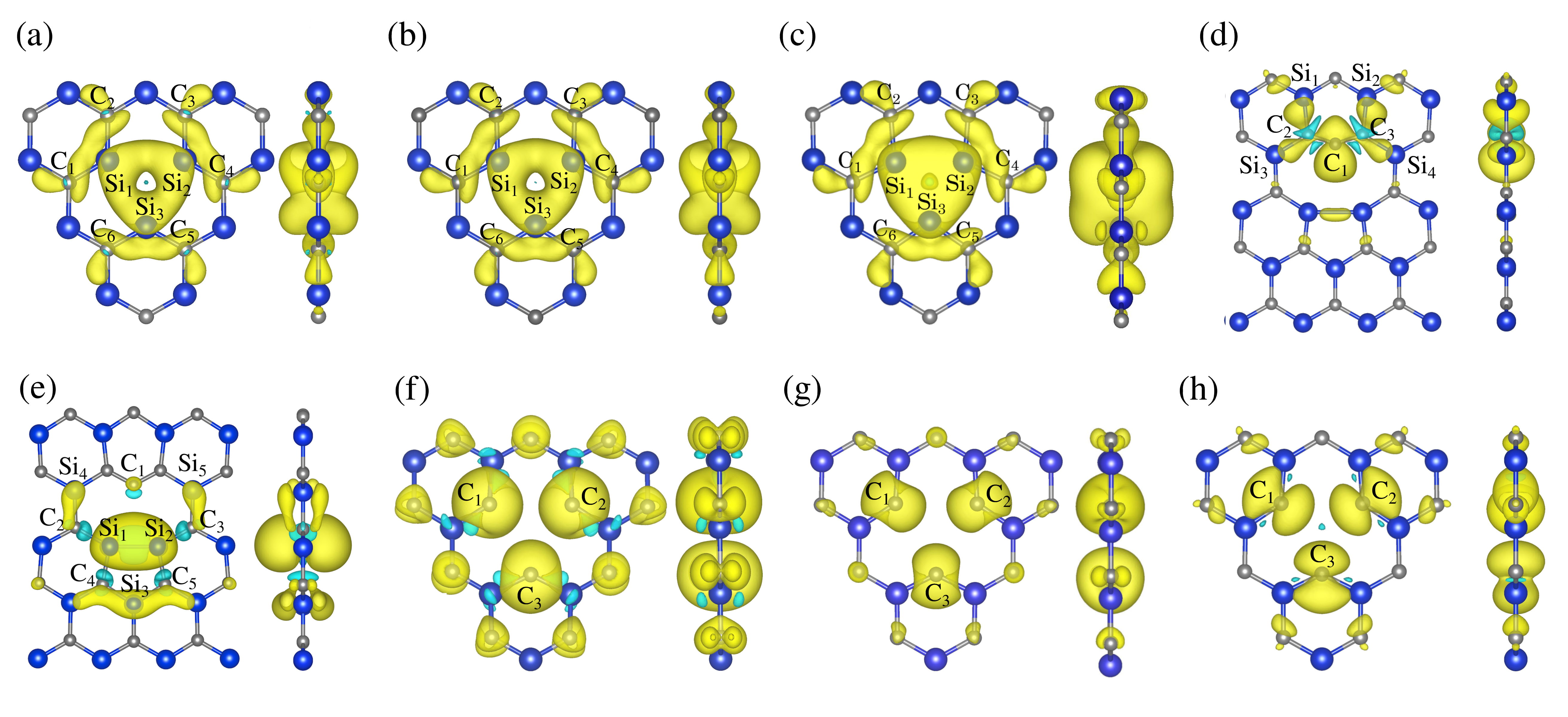}
\caption{ \label{fig:spindens}
The spin density of the (a) V$_{\text{C}}^{+}$, (b) V$_{\text{C}}^{0}$, (c) V$_{\text{C}}^{-}$, (d) V$_{\text{C}}$-C$_{\text{Si}}^{+}$, (e) V$_{\text{C}}$-C$_{\text{Si}}^{-}$, (f) V$_{\text{Si}}^{+}$, (g) V$_{\text{Si}}^{0}$, and (h) V$_{\text{Si}}^{-}$ defects from the top and side view done by HSE06 calculations. Si and C atoms are shown with blue and gray spheres, respectively. The hyperfine constants for the atoms labeled here are given in Table~\ref{tab:hf}. Light blue and yellow lobes exhibit negative and positive isovalues for the average spin density, respectively. The isosurface absolute value is $10^{-3}$~\AA$^{-3}$.
}
\end{figure*}

\begin{table}[t]
\setlength\extrarowheight{5pt}
\setlength\tabcolsep{1pt}
\caption{\label{tab:hf} 
Computed hyperfine (HF) principal values ($A_{xx}$, $A_{yy}$ and $A_{zz}$) for the nearest-neighbor $^{29}$Si and $^{13}$C $I=1/2$ isotopes by using HSE06 for the considered vacancy defects. The atom labels are shown in Fig.~\ref{fig:spindens}. Core spin polarization correction is involved in the listed HF constants. We note that the numerical accuracy of our data is about $\pm 0.3$~MHz. Also the HF constants of V$_{\text{C}}^{+}$ and V$_{\text{Si}}$ in its all charge states are less accurate due to the high correlation nature of their electronic structure. 
}
\begin{ruledtabular}
\begin{tabular}{ccccc}
Defect, spin&Atoms & $A_{xx}$ (MHz) & $A_{yy}$ (MHz) & $A_{zz}$ (MHz)  \\ \hline
V$_{\text{C}}^{+}$, 1/2&C$_1$-C$_2$$\dots$C$_6$ & 38.4&35.3 &49.0\\
&Si$_1$-Si$_2$-Si$_3$  &$-341.3$&$-329.2$&$-404.8$\\
&&&&\\
V$_{\text{C}}^{0}$, 1&C$_1$-C$_2$$\dots$C$_6$ &$41.0$& $38.3$& $51.4$\\
&Si$_1$-Si$_2$-Si$_3$  &$-380.6$ &$-370.4$ &$-437.7$ \\
&&&&\\
V$_{\text{C}}^{-}$, 3/2&C$_1$-C$_2$$\dots$C$_6$ &$29.3$& $27.8$& $36.24$\\
&Si$_1$-Si$_2$-Si$_3$  &$-289.0$ &$-265.1$ &$-300.1$ \\
&&&&\\
   V$_{\text{C}}$-C$_{\text{Si}}^{+}$, 1/2&C$_1$          & $155.5$ & $141.3$ &   $342.5$\\
    & C$_2$-C$_3$    & $-5.5$ &$3.9$ & $16.2$   \\
    & Si$_1$-Si$_2$    & $-121.3$ & $-116.8$& $-145.8$   \\
    & Si$_3$-Si$_4$    & $-9.9$ & $-9.7$ & $-12.8$    \\
 &&&&\\
  V$_{\text{C}}$-C$_{\text{Si}}^{-}$, 1/2&C$_1$          & $-2.9$ & $-7.3$ &   $-1.7$\\
  & C$_2$-C$_3$    & $-5.5$ & $-3.9$ &   $-5.9$  \\
    &   C$_4$-C$_5$    & $-5.5$ & $-3.5$ &   $-6.2$ \\
     &          Si$_4$-Si$_5$  & $-4.3$ & $-3.9$ &  $-15.7$ \\
    &                   Si$_1$-Si$_2$  & $1.5$ &  $4.5$ & $-101.3$  \\
&                   Si$_3$         & $-2.5$ & $-2.5$ &  $-26.1$  \\
&&&&\\
V$_{\text{Si}}^{+}$, 3/2 & C$_1$-C$_2$-C$_3$  &122.5 & 116.1 &200.5 \\
&&&&\\
V$_{\text{Si}}^{0}$, 1&C$_1$-C$_2$-C$_3$  &86.3 &56.3 &97.7 \\
&&&&\\
V$_{\text{Si}}^{-}$, 1/2&C$_1$-C$_2$-C$_3$  & 127.1&123.6 &204.3\\
 
\end{tabular}
\end{ruledtabular}
\end{table}

Most of the considered vacancy defects have a paramagnetic ground state or low-energy metastable state. The respective spin density distributions are displayed in Fig.~\ref{fig:spindens}. These defects may be observed by electron spin resonance techniques. 2D-SiC consists of $^{13}$C and $^{29}$Si $I=1/2$ nuclear spins with about 1.1\% and 4.5\% natural abundance; a hyperfine interaction may be developed between the electron spin and the nuclear spins that are randomly distributed in the 2D-SiC crystal. These hyperfine interactions provide an unique pattern in their electron spin resonance spectrum that could be an important fingerprint for identification. The most important atomic positions around the vacancies are labeled in Fig.~\ref{fig:spindens}. The hyperfine interaction can be fully described by the hyperfine (HF) tensor as explained in Sec.~\ref{sec:method}. The calculated HF tensor can be diagonalized yielding the three principal values, also known as the HF constants. The HF constants are labeled as $A_{xx}$, $A_{yy}$, and $A_{zz}$, and following standard convention the largest absolute value is associated with $A_{zz}$. We find that the spin density of V$_{\text{Si}}^{+}$, V$_{\text{Si}}^{0}$, V$_{\text{Si}}^{-}$ and V$_{\text{C}}^{+}$ shows symmetry breaking solution which is a sign of multi-determinant nature in the ground state. We reiterate here that to estimate the hyperfine and zero field interactions for these cases, we symmetrized the Kohn-Sham wavefunctions so the spin density according to the point group symmetry of the geometry as depicted in Fig.~\ref{fig:spindens}.


In V$_{\text{C}}^{0}$, the spin density is predominantly localized on the nearest Si atoms labeled as Si$_1$, Si$_2$, and Si$_3$ as expected because the dangling bonds of these three Si-atoms produce the orbitals in the band gap. The average HF constant for these three $^{29}$Si isotopes is $396$~MHz. 
The V$_{\text{C}}^{-}$ and V$_{\text{C}}^{+}$ defects produce very similar spin density distribution to the neutral one with the average HF constant of $285$~MHz and $358$~MHz, respectively.

In the V$_{\text{C}}$-C$_{\text{Si}}^{-}$ defect, the spin density is mainly localized on the Si$_1$ and Si$_2$ atoms, leading to an average HF constant of about $33$~MHz. On the other hand V$_{\text{C}}$-C$_{\text{Si}}^{+}$ defect has mostly localized spin density on the Si$_1$ and Si$_2$ atoms with an average HF constant of about $128$~MHz and at C$_1$ atom it is about $213$~MHz. 

In V$_{\text{Si}}$ defect, the carbon dangling bonds introduce the spin densities. As a consequence, the spin density is mostly localized on C$_1$, C$_2$, and C$_3$ ions for V$_{\text{Si}}^{0}$, V$_{\text{Si}}^{-}$ and V$_{\text{Si}}^{+}$ defects (Table~\ref{tab:hf} and Fig.~\ref{fig:spindens}) with an average HF constant of $80$~MHz, $152$~MHz and $146.4$~MHz, respectively. 

Next, we calculate the ZFS parameters for high-spin defects. For V$_\text{C}^{0}$ with $S=1$ spin state, the computed $D\approx0.1$~MHz and $E\approx0$~MHz (cf. Eq.~\eqref{eq:DE}) are surprising for planar defect structures. This can be explained by the large out-of-plane extension of the spin density creating a sphere-like shape, for which the dipolar spin-spin related ZFS vanishes. In some cases, this effect is so enhanced, that the extent of the spin density is more out-of-plane than in-plane, resulting in a negative $D$ value. As a consequence, the V$_\text{C}^{-}$ with $S=3/2$ yields $D=-408.1$~MHz and $E=32.9$~MHz. This phenomenon is only present for the silicon dangling bonds, in line with the preferential sp$^3$ hybridization of the silicon atoms. For V$_\text{Si}^{0}$ ($S=1$) and V$_\text{Si}^{+}$ ($S=3/2$), $D\approx -296.2$~MHz and  $D=4.24$~GHz are significant and characteristic, although, they may belong to the metastable state. We reiterate that we forced the high symmetry $D_{3h}$ solution for the Kohn-Sham wavefunctions in order to estimate the ZFS parameters for this defect in its $S=1$ state. We note that the large $D$ value for  V$_\text{Si}^{+}$ occurs due to the larger out-of-plane extension of the spin density, in contrast to that of V$_\text{Si}^{0}$ as depicted in Figs.~\ref{fig:spindens}(f) and (g), respectively.

The calculated HF constants for the immediate neighbor atoms of all the three defects are summarized in Table \ref{tab:hf}. We find that these defects produce very different EPR spectra and can be well identified by electron spin resonance techniques. 

\subsection{Optical properties: zero-phonon line and phonon sideband of the optical spectrum and the radiative lifetime}
\begin{figure*}[t]
\centering
\includegraphics*[width=1\textwidth]{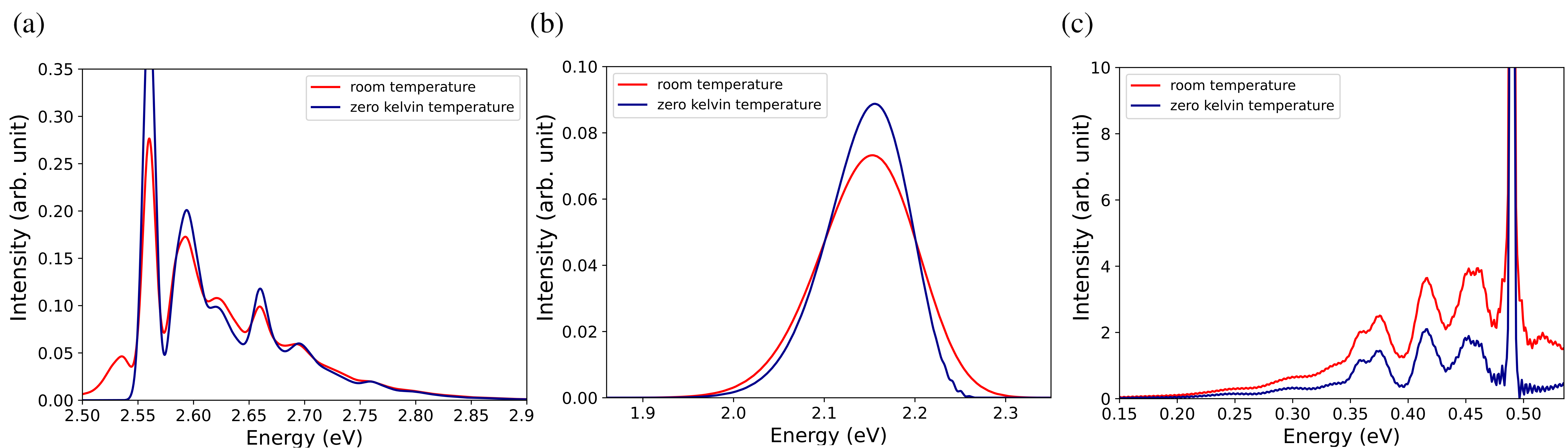}
\caption{
The optical spectrum of selected vacancy defects in 2D-SiC at zero kelvin and room temperatures. (a) Simulated absorption of V$_{\text{C}}^{0}$, (b) photoluminescence spectrum of V$_{\text{C}}^{0}$ and (c) photoluminescence spectrum of V$_{\text{C}}$-C$_{\text{Si}}^{-}$. The simulated absorption spectrum involves optical transitions to two nearby electronic excited states, see text for further details.
}
\label{fig:PLspectrum}
\end{figure*}
\begin{table*}[t]
\setlength\extrarowheight{6pt}
\setlength\tabcolsep{1pt}
\caption{\label{tab:ZPL}
Calculated zero-phonon-line energy (ZPL), ZPL wavelength ($\lambda$) by using HSE06 hybrid functional together with the calculated Huang-Rhys factor (HR) and Debye-Waller factor (DW) factor for the selected vacancy defect as obtained in the 128-atom supercell. The two lowest-energy optically allowed excited state solutions are shown for V$_{\text{C}}$-C$_{\text{Si}}^{0}$ and V$_{\text{C}}^{0}$.
}
\begin{ruledtabular}
\begin{tabular}{ c  c  c c c c c c }
Defect & Transition &  $E_{\rm ZPL}$ (eV)& $\lambda_{\rm ZPL}$ (nm)& HR & DW (\%) & $\Delta Q$ & El.\ conf.\ \\ 
\hline
V$_{\text{C}}$-C$_{\text{Si}}^{0}$&$A_{\text 1} - B_{\text 2}^{(1)}$ &0.50&2479 &8.8&0.01\% & 1.11 &$a_{\text 1} - b_{\text 2}^{(1)}$\\
& $A_{\text 1} - B_{\text 2}^{(2)}$&1.27&976  &9.41 &0.008\%  & 1.17 &$a_{\text 1} - b_{\text 2}^{(2)}$\\ 

V$_{\text{C}}$-C$_{\text{Si}}^{-}$
&$B_{\text 2}^{(1)} - B_{\text 2}^{(2)}$&0.59& 2101 &  0.80  & 44\% & 0.39 &$b_{\text 2}^{(1)} - b_{\text 2}^{(2)}$\\

V$_{\text{C}}^{0}$ &$E^\prime-E^\prime$  & 2.54 & 629 & 1.88 & 15.26\% & 0.57&$e^\prime-\text{CBM}(e^\prime)$\\
& $E^\prime-A_2^{\prime\prime}$& 2.26 & 729 & 6.57 & 0.001\% & 1.47 &  $\text{VBM}(a_1^\prime)-a_2^{\prime\prime}$\\
\end{tabular}
\end{ruledtabular}
\end{table*}

Vacancy defects introduce multiple filled and empty defect levels in the fundamental band gap in 2D-SiC. This may lead to visible or near-infrared emission upon photoexcitation below the photoionization threshold energies for the optically allowed transitions. Some of the excited states may have a zero optical transition dipole moment ($\mu$) towards the electronic ground state by symmetry that we call dark states. Although, dark states could fluoresce with the assistance of phonons, the so-called Herzberg-Teller optical transition, we ignore this scenario because the direct non-radiative decay by phonons (internal conversion) could be very competitive, in particular, for the cases where the energy gap between the ground and excited states is less then 0.5~eV.

We find that V$_\text{Si}$ defect has a highly correlated electronic structure with very complicated ground and excited states. Here the predictive power of the Kohn-Sham DFT method is questionable so we do not attempt to produce an optical signature for these defects. Nevertheless, we note that the defect is most likely stable in its negative charge state when donors are present in 2D-SiC or can be driven to this charge state by photoexcitation with laser energy at $>1.6$~eV. We can estimate the ZPL energy for V$_\text{Si}^{-}$ from the calculated $(0|-)$ occupation level with respect to the conduction band edge which yields $\approx 2$~eV. The photoluminescence spectrum should show a broad emission with a maximum intensity at lower energy than the ZPL energy coming from the phonon-assisted optical transitions because of the large relaxation energy going from the neutral to the negative charge state of the defect.   

The most stable vacancy defect is V$_{\text{C}}$, so we analyze its optical activity in detail. In Section~\ref{ssec:formen} we already discussed that high-energy illumination may dynamically drive the system into its neutral charge state irrespective of its initial charge state before illumination. Thus, we first discuss the possible optical transitions in the neutral charge state. The V$_{\text{C}}$ defect has a high $D_{3h}$ symmetry with occupied $e^\prime$ and unoccupied $a_2^{\prime\prime}$ levels in the spin majority channel as depicted in Fig.~\ref{fig:elstr}. The multiplet ground state transforms as $^3A_2^\prime$ whereas the excited state with promoting an electron from $e^\prime$ to $a_2^{\prime\prime}$ transforms as $^3E^{\prime\prime}$. The $^3A_2^\prime \rightarrow ^3E^{\prime\prime}$ is an optically forbidden transition. The calculated total energy difference is about 0.2~eV, so this dark excited state is not considered further. The optically allowed transitions come with promoting the electron from/to the defect level to/from the band edges as given in Table~\ref{tab:ZPL}. The optical transitions towards the valence band edge (VBM-transition) and the conduction band edge (CBM-transition) are significantly different. The calculated $\Delta Q$ for VBM-transition is larger than 1.0 which implies a very small DW factor so a very broad optical signal with featureless Gaussian lineshape. However, this value is only 0.57 for CBM-transition which should result in an intense ZPL peak and structured phonon sideband in the optical spectrum. We also find that the calculate $\mu$ is about $35\times$ larger for the CBM-transition than that for the VBM-transition. The absorption spectrum of the defect may be considered as the weighted sum of the two calculated absorption spectra where the relative intensity of the CBM-transition absorption spectrum is $35\times$ larger than that of the VBM-transition. The final absorption spectrum is plotted in Fig.~\ref{fig:PLspectrum}(a). The peaks in the spectrum are originated from the phonon participation in the CBM-transition. In simulating the PL spectrum, we consider the Kasha-rule which assumes that the emission comes from the lowest energy bright state. We think that the Kasha-rule applies here as the calculated energy difference between the two excited states is $\approx0.2$~eV which can be effectively bridged by phonons. So the PL of V$_\text{C}^{0}$ is determined by the VBM-transition which results in a broad PL spectrum in the visible as plotted in Fig.~\ref{fig:PLspectrum}(b). The absorption spectrum and the PL spectrum of the defect are very different because of the aforementioned reasons. The photon can likely be effectively absorbed in the CBM-transition to generate the PL spectrum in the VBM-transition in the PL process. This PL signal belongs to the triplet state of the defect. Singlet dark states also exist that were not discussed here which may contribute to a spin-selective non-radiative decay from the excited state towards the ground state via intersystem crossing. It is beyond the scope of this study to fully explore this phenomenon. Nevertheless, we speculate here that V$_\text{C}^{0}$ might be used as a quantum sensor if the spin state can be optically initialized and read out.
The radiative lifetime ($\tau$) of the photoluminescence can be given~\cite{Weisskopf1930, Thiering2018} as
\begin{equation}\label{eq:tau}
\frac{1}{\tau}=\frac{n\omega^3|\mu|^2}{3\pi \epsilon_0\hbar c^3}\text{,}
\end{equation}
where $\hbar \omega$ is the excitation energy (ZPL energy in our case), $n$ is the refractive index which is about $2.65$, $c$ is the speed of light, and $\epsilon _0$ is the vacuum permittivity. The final result is $\tau = 255$~ns which is not very bright. So it is likely that the defect can be observed in ensembles that can be harnessed in quantum sensor applications. If 2D-SiC is n-type doped and the photoexcitation energy is below the photoionization threshold then V$_\text{C}^{-}$ may be optically excited. We find that the ZPL is at 1.05~eV for the lowest energy optically allowed excited state and the calculated DW factor is extremely low which results in $\approx 1$~eV broadening with a maximum intensity at $\approx 0.5$~eV. This indicates that the non-radiative decay is a very competitive process, and the emission may be too dim to be observed.

Next, we turn to the discussion of the V$_{\text{C}}$-C$_{\text{Si}}$ defect. The V$_{\text{C}}$-C$_{\text{Si}}$ defect has $C_{2v}$ symmetry which may result in optically forbidden transitions between the $B_2$ and $B_1$ states. In V$_{\text{C}}$-C$_{\text{Si}}^{0}$, the two lowest energy excited states have nonzero optical transition dipole moment separated by 0.77~eV (see Table~\ref{tab:ZPL}). We speculate that if the excitation energy is above the second excited state's energy then the Kasha rule may break down because of the large energy gap between the second and first excited states, and emission is viable from the second excited state. Therefore, we provide the optical parameters for both excited states in Table~\ref{tab:ZPL}. The HR factors for the neutral defect are prominent ($>8$), which means a minor contribution from the coherent ZPL. We conclude that they are not applicable as quantum emitters for quantum communication applications.  

In V$_{\text{C}}$-C$_{\text{Si}}^{-}$, the energy gap between the
first and second excited states are only about 0.2~eV; thus, the Kasha rule is applied, and
the phonons can rapidly cool the system from the second excited state to the lowest-energy
excited state. Therefore, we provide the optical parameters only for the first excited state of V$_{\text{C}}$-C$_{\text{Si}}^-$. Here, the calculated HR factor is much smaller than that for V$_{\text{C}}$-C$_{\text{Si}}^0$. As a consequence, the photoluminescence spectrum displays a sharp ZPL emission even at room temperature [see Fig.~\ref{fig:PLspectrum}(c)]. The calculated ZPL energy is at around 0.6~eV which falls in the near-infrared region. Although the emission wavelength is longer than the ultralong wavelength region in commercial optical fibers, the transmission at the calculated ZPL wavelength is still effective in the optical fibers. The calculated radiative lifetime is $\tau=155$~ns.  which makes it possible to observe it at single defect level. The extraction of the emitted photon from a 2D material should be relatively efficient; therefore, the calculated optical lifetime implies that the photoluminescence intensity would be sufficient to be observed at a single defect level. Thus, we propose that V$_{\text{C}}$-C$_{\text{Si}}^{-}$ is a candidate for a solid-state quantum emitter in the near-infrared wavelength region.

We note here that optical signals were detected in 2D-SiC nanoflakes~\cite{Chabi2021}. We mention again that they found two high (2.2 and 2.6~eV) and one low (2.3~eV) intensity peaks in the absorption spectrum and a 1.0~eV-broad peak with a maximum at 2.58~eV in the PL spectrum~\cite{Chabi2021}. The 2.58-eV PL peak was associated with the band edges optical transition based on the argument that it is common with the 2.6-eV absorption peak~\cite{Chabi2021}. However, this speculation may be questioned. The 2.6-eV absorption peak has a much smaller band width ($\approx 0.2$~eV) than the 2.58-eV emission ($\approx 1.0$~eV). This rather suggests that their origin are distinct. The 1.0~eV broadening of the PL center with Gaussian lineshape indicates the phonon participation in the optical transition; the DW factor should be extremely small in this case that we can learn from the simulated PL spectrum of V$_\text{C}^0$. It is very unlikely that this PL center originates from band edges optical transition. We show that the considered vacancy defects produce distinct absorption and PL spectra. However, we do not find an one-to-one correspondence between the simulated optical spectra of the vacancy defects and the observed ones. Nevertheless, it can be concluded that the observed PL and absorption spectra should be defect-related. We note that the observed optical signals may come from the defective edges of the 2D-SiC nanoflakes which phenomenon is out of the scope of this study.

\section{Summary and Conclusions} \label{sec:summary} 

Point defects in 2D semiconductors and insulators are promising candidates as scalable coherent single-photon emitters and qubits for quantum technologies. In this work, we present results for vacancy defects in 2D-SiC as obtained by plane-wave supercell DFT calculations. The V$_{\text{C}}$ defect is the most stable one and it might be used as a quantum sensor with optically addressing its spin in the visible. The next stable defect,
V$_{\text{C}}$-C$_{\text{Si}}$, provides paramagnetic and fluorescent signals in the negative charge state. We found that this acceptor state may be realized in nitrogen-donor-doped 2D-SiC material. We found a favorable Debye-Waller factor ($44\%$) and a relatively short optical lifetime ($155$~ns) for V$_{\text{C}}$-C$_{\text{Si}}^-$ which makes this defect a good candidate to be considered as a single photon emitter in 2D-SiC. V$_{\text{Si}}$ defect might form with irradiation techniques. The defect may have metastable and stable paramagnetic states in their respective charge states, and we provide their hyperfine coupling parameters as a fingerprint in the EPR spectrum.  

\begin{acknowledgments}
The simulations have been partially performed using the resources provided by the Hungarian Governmental Information Technology Development Agency (KIF\"U). 
A.G.\ acknowledges the National Research, Development, and Innovation Office of Hungary Grant No.\ KKP129866 of the National Excellence Program of Quantum-Coherent Materials Project and Grant No.\ 2022-2.1.1-NL-2022-00004 of the Quantum Information National Laboratory supported by the Cultural and Innovation Ministry of Innovation of Hungary. T.A-N.\ has been supported in part by the Academy of Finland through its QTF Center of Excellence Grant No.~312298. We also acknowledge the computational resources provided by the Aalto Science-IT project. 
\end{acknowledgments}

\bibliography{2dSiC}

\end{document}